# Energy and Utility Optimization in Wireless Networks with Random Access

Amirmahdi Khodaian, Babak H. Khalaj,
Electrical Engineering Department
Sharif University of Technology
Tehran, Iran
Email: khodaian@ee.shrif.edu, khalaj@sharif.edu

*Abstract*— **Energy consumption is a main issue of concern in wireless networks. Energy minimization increases the time that networks' nodes work properly without recharging or substituting batteries. Another criterion for network performance is data transmission rate which is usually quantified by a network utility function. There exists an inherent tradeoff between these criteria and enhancing one of them can deteriorate the other one. In this paper, we consider both Network Utility Maximization (NUM) and energy minimization in a bi-criterion optimization problem. The problem is formulated for Random Access (RA) Medium Access Control (MAC) for ad-hoc networks. First, we optimize performance of the MAC and define utility as a monotonically increasing function of link throughputs. We investigate the optimal tradeoff between energy and utility in this part. In the second part, we define utility as a function of end to end rates and optimize MAC and transport layers simultaneously. We calculate optimal persistence probabilities and end-to-end rates. Finally, by means of duality theorem, we decompose the problem into smaller subproblems, which are solved at node and network layers separately. This decomposition avoids need for a central unit while sustaining benefits of layering.**

**Keywords- *Energy-utility tradeoff; persistence probability, random access; convex optimization; distributed algorithm.***

## I. INTRODUCTION

In this paper, an ad-hoc network is considered where there is no infrastructure and intermediate nodes send packets toward their destinations. Use of random access is common in such networks since random access algorithms are inherently distributed as nodes themselves decide when to access the channel [1]. The main characteristic of random access is independent node transmissions. This characteristic results in both an advantage and a disadvantage. The advantage is that there is no need for central controller and the disadvantage is possibility of collision. Collision occurs since there is no central controller and it is possible that two or more nodes transmit simultaneously and their packets collide. Such collisions result in waste of both energy and bandwidth, thus, network parameters such as persistence probability of nodes should be adjusted in order to optimize bandwidth and energy consumption.

The importance of energy efficiency in ad-hoc networks stems from multi-hop nature of the network. If nodes of an ad-hoc network run out of energy some routs may become disconnected [2], therefore the available energy of nodes should be consumed to transmit as much information as possible.

Another criterion for good performance of a network is network utility which is a function of allocated channel or rate to each node. Network Utility Maximization (NUM) has recently received much attention in the literature [5], [6], [7]. It is first proposed by Kelly [5] in order to optimize end-to-end rates of the wired networks. It is also used in optimizing transport layer of wireless networks [6],[7]. Nandagopal et. al. [8] used similar approach in proportionally fair channel allocation and [9] developed the idea of optimizing persistence probabilities in random access wireless networks and designing MAC protocols.

Energy efficiency and utility maximization are important objectives in Wireless Sensor Networks (WSN). A WSN collects information from different points of the field and it is a performance criterion for WSN to maximize information collected from all regions of the network [3]. Minimizing energy is also important in WSNs, because it is usually impossible to recharge batteries of WSN nodes and when the batteries run out of energy so do the nodes [4]. Thus, WSNs need both utility maximization and energy minimization.

In this paper, we investigate both energy minimization and NUM in a bi-criterion optimization problem. We propose distributed algorithms that can be used to tradeoff energy and utility. Energy minimization and lifetime maximization for wireless ad-hoc networks have been the focal point of many research activities [10], [11]. However to the best knowledge of the authors our work is the first one which considers energy minimization in random access networks. [6] and [9] have formulated and solved NUM for random access but they have not considered energy consumption. Tradeoff between utility and network lifetime is investigated in [12] but it has not considered random access as well.

The rest of paper is organized as follows. In the next section network model is presented. Then, in section III we concentrate on MAC optimization and define utility as a function of link throughputs. Tradeoff between utility and energy consumption is also found. Cross layer optimization of MAC and transport layer is described in section IV where we optimize both layers in order to minimize energy and provide

This work is supported in part by Advanced Communication Research Institute (ACRI).

maximum utility in transport layer. Section V contains numerical results and discusses advantages of cross layer optimization. We conclude paper and review its contributions in section VI.

## II. NETWORK MODEL

Suppose a set of nodes, *N*, want to transmit their packets through their neighbors using the set of links *L*. Each node selects one of its links and transmits with probability $p_{ij}$ where *i* is transmitter index and *j* is receiver index. We show the transmission probability of node *i* with $P_i$, which is summation of persistent probabilities of output links. The set of nodes that receive power from node *i* is shown by $N_i^{out}$ and the set of nodes that *i* receive power from them is shown by $N_i^{in}$. It is evident that if nodes hear each other symmetrically, then $N_i^{out} = N_i^{in}$. However, this assumption is not a presupposition in this paper. We also denote the set of nodes which *i* transmits to them with $O_i$ and the set of nodes that transmit to *i* with $I_i$. We define connectivity factor by ratio of communication range to network dimension. Thus as the node's power increases connectivity factor and number of out neighbors, $|N_i^{out}|$, will increase.

Network links are used by set *S* of information sources. Source $s \in S$ uses subset of links, $L(s) \in L$, as a route to transmit its data. The set of sources that share link (*i*, *j*) is defined by $S(i,j) = \{s \in S | (i,j) \in L(s)\}$ and therefore $(i,j) \in L(s)$ if and only if $s \in S(i,j)$. We suppose that link (*i*,*j*) has a fixed capacity of $c_{ij}$ and transmission rate of sources *s* is $y_s$.

## III. OPTIMIZATION OF MAC

In this section we optimize persistent probabilities in order to maximize network utility function and minimize energy consumption. A common method for solving such a bi-criterion optimization problem and achieving Pareto optimal points is scalarization[13]. Using this method, we set objective function as linear combination of energy and utility with $\lambda_1 > 0$ and $-\lambda_2 < 0$ as coefficients:

$$\begin{aligned}
\min \quad & f = \lambda_1 E - \lambda_2 U \\
S.t. \quad & 0 \leq P_i \leq 1 & \forall i \in N \\
& 0 \leq p_{ij} \leq 1 & \forall i \in N, j \in O_i \\
& P_i = \sum_{j \in O_i} p_{ij} & \forall i \in N
\end{aligned} \quad (1)$$

A negative coefficient is used for utility function since minimizing –*U* is equivalent to maximizing *U*. The constraints of (1) ensure that optimal persistent probabilities have valid values and summation of persistent probabilities of output links of each node is equal to transmission probability of the node.

Utility function, *U*, is defined as summation of link utilities and to achieve a proportional fairness between links we use the same approach as [8] and [9]. We define utility as a logarithmic function of the link throughput.

$$U = \sum_{(i,j) \in L} \log(x_{ij}) \quad (2)$$

In order to calculate throughput of the links, $x_{ij}$, we suppose that successful packet reception of each node depends only on transmission of its in-neighbors. Therefore, a packet is received successfully if and only if neither the receiver nor any of the receiver in-neighbors except the transmitter have sent packets in the same time. Thus, throughput of a link is multiplication of successful reception probability and link capacity, and is given by:

$$x_{ij} = c_{ij} p_{ij} (1 - P_j) \prod_{l \in N_j^{in} - \{i\}} (1 - P_l) \quad (3)$$

If the required energy to transmit a packet by node *i* is equal to $e_i$, average energy consumption of a node in one timeslot is given by $E_i = e_i \times P_i$. Thus total energy consumption of the network is:

$$E = \sum_{i \in N} E_i = \sum_{i \in N} e_i P_i \quad (4)$$

Applying (2)-(4) in (1) and reordering terms we have:

$$\lambda_1 E - \lambda_2 U = \sum_{i \in N} \lambda_1 e_i P_i \\
- \lambda_2 \sum_{i \in N} \sum_{j \in O_i} \log(c_{ij} p_{ij}) + [\sum_{k \in N_i^{out}} |I_k| + |I_i| - |O_i|] \log(1 - P_i) \quad (5)$$

For further simplifications we prove the following lemma.

**Lemma 1:** Optimal link and node probabilities are related to each other by $p_{ij}^* = P_i^* / |O_i|$ ; $i \in N, j \in O_i$.

**Proof**: First we show that optimal link probabilities of node *i*, should be equal to each other. Suppose $p_{ij}^* \neq p_{il}^*$ ; $j, l \in O_i$. If we replace both of these probabilities with $(p_{ij}^* + p_{il}^*)/2$ this will not affect $P_i$ and therefore it will neither change *E* nor violate the constraints. This conversion can increase utility because we have:

$$(p_{ij}^* - p_{il}^*)^2 > 0 \Rightarrow (p_{ij}^* + p_{il}^*)^2 > 4 p_{ij}^* p_{il}^* \\
\Rightarrow 2 \log(\frac{p_{ij}^* + p_{il}^*}{2}) > \log(p_{ij}^*) + \log(p_{il}^*) \quad (6)$$

Thus, $p_{ij}^* = p_{il}^* ; \forall j, l \in O_i$ and since number of output links of *i* is equal to $|O_i|$ we have: $p_{ij}^* = P_i^* / |O_i|$ ; $i \in N, j \in O_i$.

If we apply Lemma 1 in (5) then utility will become a function of node transmission probabilities. With some algebraic manipulations first and second derivative of utility function with respect to $P_i$ are given by:

$$f'_{P_i} = \frac{\partial f}{\partial P_i} = A_i - \frac{B_i}{P_i} + \frac{C_i}{1 - P_i} \qquad \forall i \in N \quad (7)$$

$$f''_{P_i} = \frac{\partial^2 f}{\partial P_i^2} = \frac{B_i}{P_i^2} + \frac{C_i}{(1 - P_i)^2} \qquad \forall i \in N \quad (8)$$

where: $A_i = \lambda_1 e_i > 0$, $B_i = \lambda_2 |O_i| > 0$, $C_i = \lambda_2 [\sum_{k \in N_i^{out}} |I_k| + |I_i| - |O_i|] \geq 0$. $C_i = 0$ iff only *i* transmits to node j and its neighbors. In this special case, which we ignore hereafter, $P_i^* = min(1, B_i/A_i)$.

Therefore, $f''_{P_i} > 0$ and $\nabla^2_{\mathbf{P}} f$ is positive definite. Thus, problem (1) is a convex optimization problem and it has a unique solution which is the stationary point of problem and can be found by setting (7) equal to zero. According to the the following Lemma, this stationary point satisfies the constraints.

**Lemma 2:** $f'_{P_i} = 0$ has a unique solution in interval (0, 1).

**Proof:** With respect to (7) $P_i \to 0^+ \Rightarrow f'_{P_i} \to -\infty$ and $P_i \to 1^- \Rightarrow f'_{P_i} \to +\infty$. Since $f'_{P_i}$ is continuous and ascending function of $P_i$ it will become zero only in one point of interval (0, 1).

If values of $\lambda_1$ and $\lambda_2$ are specified then each node can solve $f'_{P_i} = 0$ and achieve optimal solution. In order to do so, each node requires local information such as number of its incoming and outgoing links, and number of incoming links of out-neighbors. It should be noted that $\lambda_1$ and $\lambda_2$ can be used to tradeoff energy and utility. For example, if we set $\lambda_1=0$ then we have ignored energy and the problem becomes utility maximization with solution: $P_i = \dfrac{|O_i|}{\sum_{k \in N_i^{out}} |I_k| + |I_i|}$. Also if we set $\lambda_2=0$, the trivial solution of $P_i=0$ will be achieved.

## IV. CROSS-LAYER OPTIMIZATION OF MAC AND TRANSPORT

In this section we define network utility as a function of end-to-end rates, *i.e.* $U = \sum_{s \in S(i,j)} \log(y_s)$. The bi-criterion problem of utility maximization and energy minimization can be formulated as follows:

$$\lambda_1 \sum_{i \in N} P_i e_i - \lambda_2 \sum_{s \in S} \log(y_s)$$

$$\sum_{S(i,j)} y_s \leq x_{ij} \quad \forall (i,j) \in L$$

$$\sum_{j \in O_i} p_{ij} = P_i \quad \forall i \in N \quad (9)$$

$$0 \leq p_{ij} \leq 1 \quad \forall (i,j) \in L$$

$$P_i \leq 1 \quad \forall i \in N \ ; \ y_s \geq 0 \quad \forall s \in S$$

The first constraint limits the summation of source rates which pass through link $(i,j)$ to be less than link throughput. Remaining constraints ensure validity of link and node transmission probabilities. In this problem, the objective function is a convex function of node transmission probabilities and source rates and constraints 2 and 3 are linear. However, the first constraint is a nonlinear and non-convex function. In order to formulate the problem as a convex problem, we change variables to $z_s = \log(y_s)$ and apply logarithmic function to the first constraint (It is obvious that such transformation does not affect constraint since $\log(\cdot)$ is an ascending function).

$$\min_{\mathbf{p},z} \lambda_1 \sum_{i \in N} P_i e_i - \lambda_2 \sum_{s \in S} z_s$$

$$s.t. \ \log \sum_{s \in S(i,j)} e^{z_s} \leq \log(x_{ij}) \quad ; \forall (i,j) \in L \quad (10)$$

$$\textit{Transmission probability constraints}$$

It is shown in [13] that $\log \sum_i e^{\delta_i}$ is a convex function of $\delta_i$. Also $\log(x_i)$ can be computed using (3) and is a logarithmic function of link transmission probabilities. Therefore, (10) is a convex problem and we can achieve its global optimum using algorithms such as interior point method or Sequential Quadratic Programming (SQP)[14]. However these algorithms require a central unit which collects information from network's topology, solves (10) and finally sends results to the nodes. In the next section, we propose a distributed solution in which nodes achieve optimal point in an iterative process.

### A. A Distibuted Algorithm

We use the dual decomposition approach to obtain a distributed algorithm. First, we write down the Lagrangian function associated with problem (9), where $\mu_{ij}$ is Lagrange multiplier on link *l*:

$$L(\mathbf{p}, \mathbf{y}, \boldsymbol{\mu}) = \lambda_1 \sum_{i \in N} P_i e_i$$
$$- \lambda_2 \sum_{s \in S} \log(y_s) - \sum_{(i,j) \in l} \mu_{ij} (x_{ij}(\mathbf{p}) - \sum_{t \in S(i,j)} y_t) \quad (11)$$

Note that in this Lagrangian we do not relax transmission probability constraints. The Lagrange dual function can be decomposed into two parts:

$$D_1(\boldsymbol{\mu}) = \min_{y_s} \ \lambda_1 \sum_{i \in N} p_i e_i - \sum_{(i,j) \in L} \mu_{ij} x_{ij}(\mathbf{p})$$
$$s.t. \ 0 \leq p_{ij} \leq 1, \ P_i \prec 1, \ \sum_{i \in o_i} p_{ij} = P_i \quad (12)$$

$$D_2(\boldsymbol{\mu}) = \min_{y_s} \ -\lambda_2 \sum_{s \in S} \log(y_s) + \sum_{(i,j) \in L} \mu_{ij} \sum_{t \in S(i,j)} y_t$$
$$s.t. \ y_s \geq 0 \quad (13)$$

Master Dual problem is as follows:

$$\max \ D(\boldsymbol{\mu}) = D_1(\boldsymbol{\mu}) + D_2(\boldsymbol{\mu})$$
$$S.t. \ \boldsymbol{\mu} \geq 0 \quad (14)$$

It is apparent that problem (12) is a function of MAC layer parameters and (13) is a function of transport parameters. Thus, we have decomposed the dual problem into MAC and transport layer problems. Lagrangian multipliers $\mu_{ij}$, are messages that exchange information between MAC and transport layers. Since there is no compact solution for (12) we use the projected gradient method to solve it. Although gradient method is not very fast (in comparison with second order algorithms such as Newton) but its main benefit is that it require only local information. Consequently, we update $p_{ij}$ as follows:

$$^{1)} = Proj_i(p_{ij}^{(n)} - \alpha(\lambda_1 e_i - \sum_{(s,t) \in L} \mu_{st} \frac{\partial x_{st}(\mathbf{p}^{(n)})}{\partial p_{ij}})) \quad (15)$$

where $\alpha$ is optimization step and $\frac{\partial x_{(s,t)}(\underline{p}^n)}{\partial p_{ij}}$ can be calculated by (3). Also $Proj_i$ projects $p_{ij}$ on feasible region which includes two steps:

$$p_{ij}^{\mathrm{I}} = \begin{cases} p_{ij} - \frac{\sum_{j \in O_i} p_{ij} - 1}{|O_i|} & ; \sum_{j \in O_i} p_{ij} > 1 \\ p_{ij} & ; else \end{cases} \quad (16)$$

$$p_{ij}^{\mathrm{II}} = \begin{cases} 0 & ; p_{ij}^{\mathrm{I}} < 0 \\ 1 & ; p_{ij}^{\mathrm{I}} > 1 \\ p_{ij}^{\mathrm{I}} & ; else \end{cases} \quad (17)$$

In order to solve (13) first we rearrange its terms:

$$D_2(\mu) = \min_{y_s} -\lambda_2 \sum_{s \in S} \log(y_s) + \sum_{t \in S} \sum_{(i,j) \in L(t)} \mu_{ij} y_t \quad (18)$$
$$s.t. \quad y_s \geq 0$$

By setting the first derivative of $D_2$ equal to zero we get:

$$-\lambda_2 / y_s + \sum_{(i,j) \in L(s)} \mu_{ij} = 0 \; ; \; \mu^s = \sum_{(i,j) \in L(s)} \mu_{ij}$$
$$\Rightarrow y_s = \lambda_2 / \mu^s \quad (19)$$

Since $\mu^s > 0$ and $\lambda_2 > 0$, we have $y_s > 0$.

In order to find $\mu$ we should solve (14). This is similar to the approach of [7] for optimization of transport layer. Although [7] considers utility maximization, a few changes are required in our case. Thus, we apply projected gradient method and update Lagrange multipliers as follows:

$$\mu_{ij}^{(n+1)} = [\mu_{ij}^{(n)} + \gamma(y^{(i,j)}(\boldsymbol{\mu}^{(n)}) - x_{ij}^{(n)})]^+ \quad (20)$$

where $\gamma > 0$ is optimization step size, $[z]^+ = \max\{z, 0\}$ and $y^{(i,j)}(\boldsymbol{\mu}) = \sum_{s \in S(i,j)} y_s(\boldsymbol{\mu})$.

The steps of the distributed algorithm are summarized as follows:

1. Choose initial values for link transmission probabilities and Lagrangian multipliers so that $0 < p_{ij} < 1$ and $0 < \boldsymbol{\mu}$.

2. Update Lagrange multipliers of each link with (20) and send new values to the source of all sessions that use the link.

3. By use of Lagrange multipliers and (19) update source rates and send results to links of each session.

4. Update link transmission probabilities using (15)-(17) and compute link throughputs and send them to transport layer.

5. Repeat steps 2-4 until the algorithm converges.

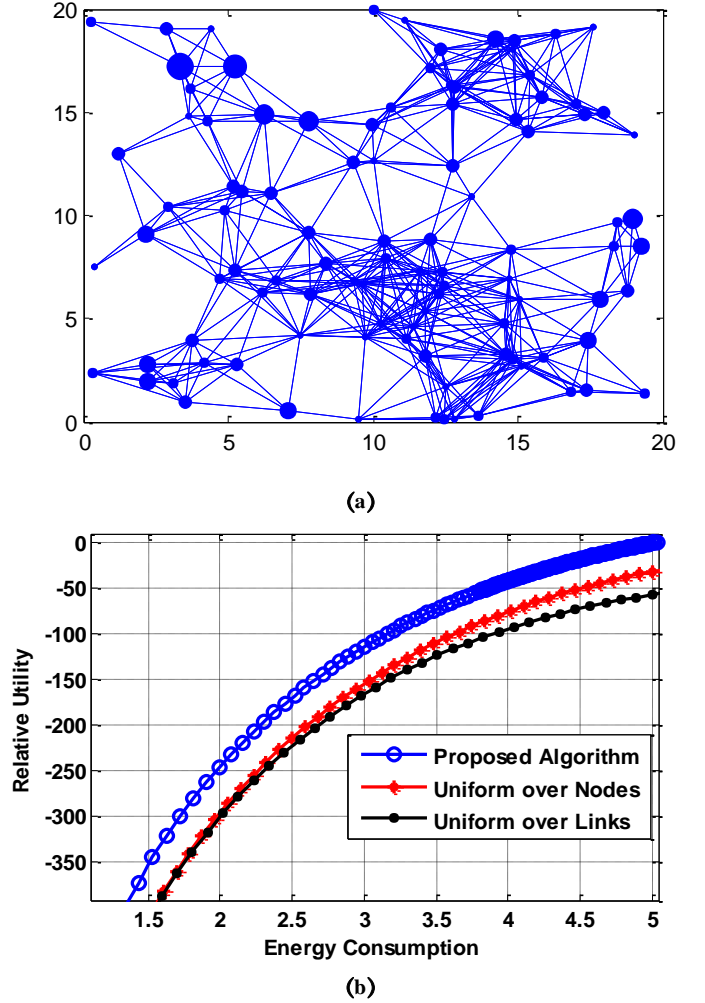

Figure 1. **MAC Optimization a) node sizes are proportional to their transmission values. b) Energy-Utility optimal tradeoff curve.**

## V. NUMERICAL RESULTS

### A. MAC Optimization

We suppose a network with 100 nodes and calculate optimal transmission probabilities with (7) for a specified value of $\lambda_1$ and $\lambda_2$. As is shown in Fig. 1-a the optimal transmission probability is small in dense regions in order to avoid packet collisions. By solving problem (1) for different values of $\lambda_1$ and $\lambda_2$ with SQP we can find the optimal tradeoff curve of energy and utility. These Pareto optimal points are sketched in Fig. 1-b where we have shown the utility relative to its maximum possible value.

In this simulation, we have supposed that the connectivity factor of each node is a random variable with uniform distribution over [0.15, 0.25]. The results are also averaged over 20 networks. Comparison of the optimal point with the uniform node transmission probability case shows that optimal solution is about 12% more energy efficient. In addition, it is about 25% more energy efficient than the uniform link transmission probability scenario.

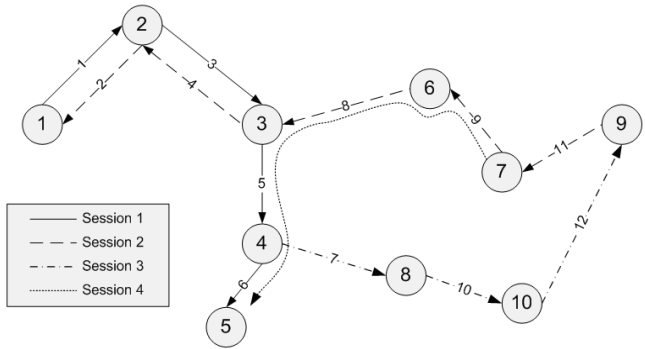

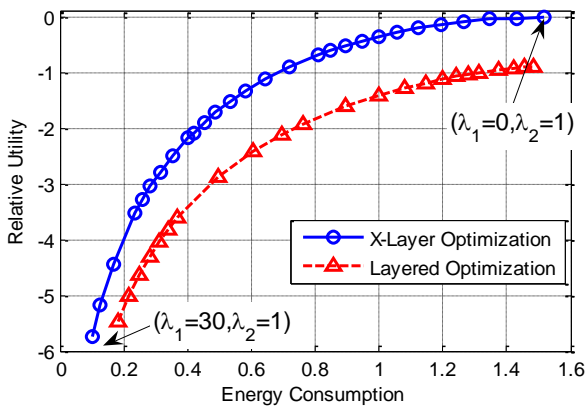

Figure 2. **Cross-layer optimization a) a sample of ad-hoc network and sessions b) Energy-utility tradeoff in cross layer optimization and comparing cross layer optimization with layer by layer optimization**

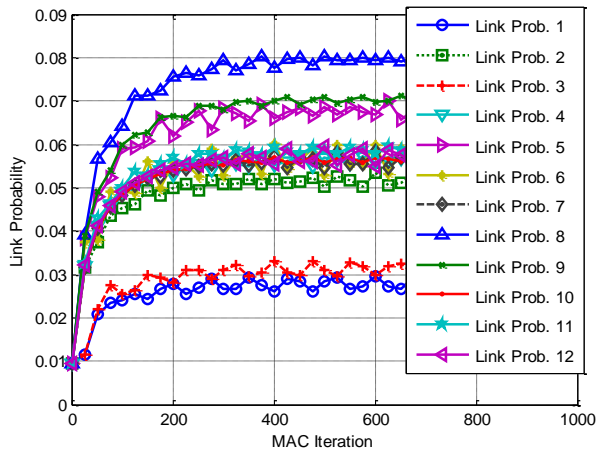

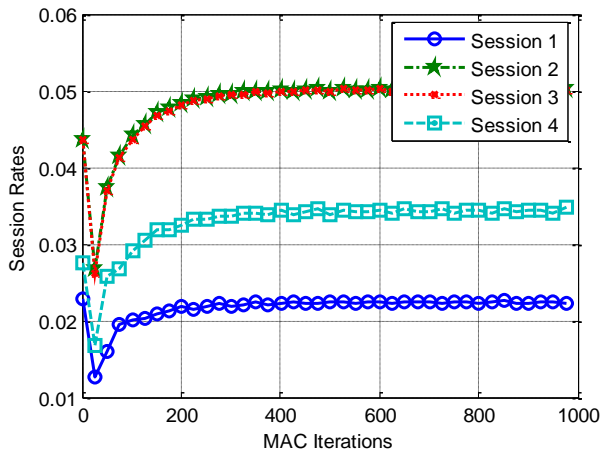

Figure 3. **Network parameters in different iterations of the distributed algorithm a) link transmission probabilities b) session rates (source rates)**

## B. MAC and transport Optimization

We consider the network shown in Fig 2(a) for cross layer optimization of MAC and transport layers and compute the optimal tradeoff curve by setting $\lambda_2=1$ and changing $\lambda_1$ over [0,30]. By setting $\lambda_1=0; \lambda_2=1$ the problem becomes a utility maximization problem. Therefore, this point shows the maximum achievable network utility. We have also compared layer by layer and cross layer optimization. In layer by layer optimization we have first computed transmission probabilities with MAC utility and energy optimization and then with the achieved throughput of links we optimize source rates. It can be seen that cross layer optimization is about 30% to 50% more energy efficient. It can also be seen that maximum achievable utility with cross layer optimization is 1 unit greater than layer by layer optimization.

We have also simulated the distributed algorithm given in section IV.A. We set $\lambda_1=5; \lambda_2=1$, transport optimization step as $\gamma_2=2$ and MAC optimization step as $\alpha=10^{-4}$. It is shown in Fig. 3 that link transmission probabilities and source rates converge after 300 iterations.

## VI. CONCLUDING REMARKS

It is shown in this paper that we can apply mathematical programming to optimize random access network parameters. We also showed that after formulating problem and making some changes, the problem can be decomposed between network layers. One of the main contributions of this paper is computing optimal tradeoff curve of Energy and utility. These curves not only can be used in network design but also specify the achievable utility and energy.

In this paper we first optimized MAC layer of wireless random access ad-hoc networks where nodes calculate optimal transmission probabilities with some local information. Energy minimization and utility maximization are special case of the solved problem. We also considered cross layer optimization of MAC and transport layers where utility is a function of source rates. We proposed a distributed algorithm that decompose problem between layers and nodes. Our numerical results show that we assign resources to links that carry a large amount of information.